\documentstyle[sprocl,12pt]{article}

\def\bea{\begin{eqnarray}}
\def\eea{\end{eqnarray}}

\begin{document}

\title{COSMIC ACCELERATION:\\ INHOMOGENEITY VERSUS VACUUM ENERGY\footnotemark}
\author{J.-F. PASCUAL-S\'ANCHEZ}
\address{Dept. Matem\'atica Aplicada Fundamental, 
Secci\'on Facultad de Ciencias,\\
Universidad de Valladolid, 47011, Valladolid,\\ SPAIN\\ 
E-mails: jfpascua@maf.uva.es; passa@gbien.tel.uva.es}

\maketitle

\footnotetext{This essay received an "Honorable Mention" in the 1999 Essay Competition of the Gravity Research Foundation.}

\abstract{In this essay, I present an alternative explanation for the cosmic acceleration which appears as a consequence of recent high redshift Supernova data. In the usual interpretation, this cosmic acceleration is explained by the presence of a positive cosmological constant or vacuum energy, in the background of Friedmann models. Instead, I will consider a Local Rotational Symmetric (LRS) inhomogeneous spacetime, with a barotropic perfect fluid equation of state for the cosmic matter. Within this framework the kinematical acceleration of the cosmic fluid or, equivalently, the inhomogeneity of matter, is just the responsible of the SNe Ia measured cosmic acceleration. Although in our model the Cosmological Principle is relaxed, it maintains local isotropy about our worldline in agreement with the CBR experiments.}

\newpage

\setlength{\parskip}{4mm}                 

\section{INTRODUCTION}
	Last year, two independent groups [1,2], 
by using type Ia Supernovae as standard candles without evolution effects, were able to extend the Hubble diagram of luminosity distance versus redshift, out to a redshift of $z \stackrel{<}{{_\sim}} 1$, implementing a generalized K-correction, [3]. 

	The main conclusion of these works is that the deceleration parameter at present cosmic time $q_0$, is negative, i.e., an acceleration of the cosmic expansion. Moreover, they interpret this conclusion in the framework of the Friedmann (FLRW) models with cosmological constant, $\Lambda$, [4], in which a necessary and sufficient condition of cosmic acceleration (regardless of the value and sign of the intrinsic spatial curvature of the cosmic spaces, if the null energy condition (NEC) holds) is that $\Lambda$ is positive.

	The cosmological constant $\Lambda$, was reinterpreted as a vacuum energy and was used by Guth [5] in the inflationary models. An estimate of this vacuum "quantum-mechanical" energy [6], is at least 120 orders of magnitude higher, that the vacuum energy associated with $\Lambda$, determined by the interpretation of the Supernova (SNe Ia) data in the background of FLRW models with $\Lambda$. Nobody knows which is the supression mechanism, if it exists.

	In this essay, I propose an alternative explanation for the measured cosmic acceleration in which, from the beginning, $\Lambda$ and hence the vacuum energy are set to zero. Our starting point will be the relaxation of the essential assumption of the FLRW models, the Cosmological Principle and, after, the consideration of barotropic locally rotational symmetric inhomogeneous models without $\Lambda$, in which the acceleration of the congruence of cosmic matter is just a sufficient condition for the SNe Ia measured cosmic acceleration.

	Our main "a priori" argument to discard the Cosmological Principle is Ockham's razor applied to observational Cosmology. 
Observationally we can only assert that there is isotropy about our worldline and this has been falsified using different tests, being the most important one, the measured high degree of isotropy of the cosmic background radiation, CBR. This local isotropy about our worldline, when combined with the Copernican Principle, leads to isotropy about all worldlines (at late times, of different clusters of Galaxies and, at early times, of the average motion of a mixture of gas and radiation)
and thus to the homogeneity of the 3-dim spacelike hypersurfaces of constant cosmic time and finally to the FLRW models.

	However, as Ellis et al. pointed out [7], if we suspend the Copernican assumption in favour of a direct observational approach, then it turns out that the local isotropy of the CBR is insufficient to force isotropy into the spacetime geometry and hence spatial homogeneity of the 3-dim cosmic hypersurfaces, i.e., to force the verification of the Cosmological Principle.

Homogeneity of the 3-dim spacelike hypersurfaces have poor observational support. At the large-scale level, we only have data from our past light cone and testing homogeneity of the 3-dim hypersurfaces at constant cosmic time, requires us to know about conditions at great distances at  present cosmic time, whereas what we can observe at great distances is what happened long time ago. So to test homogeneity of spacelike cosmic hypersurfaces,  we first have to understand how is the evolution of both the spacetime geometry and its matter-energy contents. 
For other critics to the Cosmological Principle and the FLRW models, see for instance [8,9]. 

\section{OUR MODEL: BAROTROPIC INHOMOGENEOUS LRS}
There is one family of spacetimes in which the Cosmological Principle is relaxed but they assure the observational local isotropy, these are the local rotational symmetric (LRS) inhomogeneous models. In the family of inhomogeneous LRS spacetimes,
the symmetry group
is 3-dimensional, just half the symmetry group  of the FLRW models.

In our model, I will assume that the matter part of Einstein equations have a perfect fluid form. However, we will not consider the dust case, i.e., the Lemaitre-Tolman-Bondi (LTB) models, because then necessarily the congruence of matter worldlines will be geodesic. Instead, I will consider a barotropic equation of state, $p=p(\varrho)$ and $\varrho+p>0$ (NEC condition),
 which allows for an accelerating congruence.

Geometrically, in these LRS models, the coefficients of the spacetime metric depend on two independent variables (of cosmic time and a radial coordinate), and if one chooses a comoving system, 
then the metric depends on three non-negative coefficients, and reads 	
\begin{equation}\label{1}
ds^2=-A^2(r,t)\, dt^2+B^2(r,t)\, dr^2+R^2(r,t)\, d\Omega^2 .
\end{equation}
If moreover one supposes spherical symmetry (SS), the congruence of matter fluid is initially irrotational and, then, by the supposed barotropic equation of state, the vorticity is zero at any time, where by SS,  $\varrho=\varrho(r,t)$ and $p=p(r,t)$. However, the other kinematical quantities of the congruence of matter worldlines, i.e., acceleration, shear and expansion are non zero in this spacetime. Note that in the FLRW models all are zero except the expansion.

 As the vorticity is always zero, then the fluid matter flow is always hypersurface orthogonal and there exists: 1) A cosmic time $t$
and 2) A 3-metric of the spacelike hypersurfaces. As far as I know, this spacetime was used by Mashhoon and Partovi to describe the gravitational collapse of a charged fluid sphere [10], and to large-scale observational relations [11].

From the Einstein equations without $\Lambda$, one obtains the conservation of energy-momentum $T^{ab}$:
\begin{equation}\label{2}
\nabla_aT^{ab}=0   .
\end{equation}
From (\ref{2}), one obtains (see [12]) for a perfect fluid, the energy conservation equation:
\begin{equation}\label{3}
\frac{\partial \varrho}{\partial t}+ (\varrho +p)\,\theta=0,
\end{equation}
being $\theta$ the expansion of the matter fluid and the Euler equation
\begin{equation}\label{4}
\frac{\partial p} {\partial r}+ (\varrho +p)\, \bf{a}=0,
\end{equation}
being $\bf{a}$, the acceleration of the fluid congruence.
Note that in FLRW models, equation (\ref{4}), is a tautology, because both terms on the LHS are independently zero. However,
the consequences of Euler equation (\ref{4}), are very important in our model. As the fluid is barotropic and the NEC holds, the acceleration is always away from a high-pressure region towards a neighbouring low-pressure one. In other words, the radial gradient of pressure is negative and gives place to an acceleration of the matter flow which opposes the gravitational attraction. This can also be important in order to surpass the classical singularity theorems, but in this essay, I will only show that this fluid acceleration can explain the SNe Ia data about the negativeness of the $q_0$ parameter.

\section{LUMINOSITY DISTANCE-REDSHIFT RELATION AND DECELERATION PARAMETER}
To relate our model with the SNe Ia data, we need to know how the luminosity distance-redshift relation and the deceleration parameter are modified by the inhomogeneity. By using conservation of light flux, (see [14]), it follows from the metric (\ref{1})
\begin{equation}\label{5}
D_L=(1+z)^2R(t_s,r_s),
\end{equation}
being $D_L$ the luminosity distance and $t_s, r_s$ the cosmic time and radial coordinate at emission. At present time, $t_o$, this relation reads
\begin{equation}\label{6}
D_L(t_0,z)=(1+z)^2R[t_s(t_0,z),r_s(t_0,z)].
\end{equation}
If one makes an expansion of $D_L$ to second order in $z$, after making an expansion to first order in $z$ of $t_s(t_0,z)$ and $r_s(t_0,z)$, one finds, [11]:
\begin{equation}\label{7}
D_L(t_0,z) \approx \frac{1}{H_0}[z+\textstyle\frac{1}{2}(1-Q_0)z^2]  ,
\end{equation}
where $Q_0$ is a generalized deceleration parameter at present cosmic time.
On the other hand, if one develops the metric coefficients of (\ref{1}) and the mass-energy and pressure in power series of the radial coordinate and after imposing the Einstein equations, one obtains after a scale change in the radial coordinate, [11,13]:
\begin{eqnarray}\label{8}
ds^2 \approx& &- \left(1+\textstyle\frac{1}{2}\alpha(t) r^2\right)\, dt^2                    \nonumber\\
          &  &+S^2(t)\left[\left(1+\textstyle\frac{1}{2}\beta(t) r^2\right)\, dr^2 + r^2\left(1+\textstyle\frac{1}{2}\gamma(t) r^2\right)^2 d\Omega^2\right]  ,
\end{eqnarray}
where $S(t)$ is the customary scale factor, $\alpha(t)$ is a non-negative function related to the acceleration of the cosmic fluid and a combination of $\beta$ and $\gamma$ gives the intrinsic spatial curvature of the 3-dim spacelike cosmic spaces. 

On the basis of equations (\ref{7},\ref{8}), one finds [11,13], that 
\begin{equation}\label{10}
Q_0 = q_0 - I\!\!I_0 ,
\end{equation}
thus, the luminosity distance-redshift relation at present time, reads
\begin{equation}\label{11}
D_L(t_0,z) \approx \frac{1}{H_0}[z+\textstyle\frac{1}{2}(1-q_0+I\!\!I_0)z^2] ,
\end{equation}
where $H_0$ and $q_0$ are the customary Hubble and deceleration parameters
\[H_0:=\frac{\dot{S_0}}{S_0},\]
\[q_0:=-\frac{S_0 \ddot{S_0}}{\dot{S_0}^2},\]
and $I\!\!I_0$ is a new inhomogeneity parameter which reads 
\begin{equation}\label{12}
I\!\!I_0= \frac{\alpha(t_0)}{(S_0H_0)^2}  .
\end{equation}
Note that  $I\!\!I_0$ is related to the congruence acceleration through the metric coefficient $\alpha$.
In our model the deceleration parameter at present time is:
\begin{equation}\label{13}
q_0=\textstyle\frac{1}{2}\Omega_0\left( 1 + \displaystyle\frac{3p_0}{\varrho_0}\right)- I\!\!I_0,
\end{equation}
as at present time $\displaystyle\frac{3p_0}{\varrho_0}\ll 1$, then one finally obtains
\begin {equation}\label{14}
q_0 \approx\textstyle\frac{1}{2}\Omega_0- I\!\!I_0,
\end{equation}
where $\Omega_0$ is the present matter density in units of the critical density.


\section{CONCLUSION}	
From the formula (\ref{14}), we see that one can obtain a negative deceleration parameter, i.e., cosmic acceleration, in agreement with recent SNe Ia data, by the presence of a positive inhomogeneity parameter related to the kinematic acceleration or, equivalently, to a negative pressure gradient of a cosmic barotropic fluid. In this way, it is not necessary to explain the Supernova data by the presence of $\Lambda$ or a vacuum energy or some other exotic forms of matter. Although in our model without $\Lambda$, the Cosmological Principle is relaxed, however, it maintains perfect agreement with the  local isotropy about our worldline measured by the CBR experiments.

\section{ACKNOWLEDGEMENTS}
I am grateful to P. Ruiz-Lapuente for explaining me some observational insights and to A. San Miguel and F. Vicente for discussions and TeX help. This work has been partially supported by the spanish research projects VA61/98, VA34/99 of Junta de Castilla y Le\'on and C.I.C.Y.T. PB97-0487.

\section{REFERENCES}

[1] Perlmutter, S. et al., {\it Nature}, {\bf 391} (1998) 51.\\
\noindent
[2] Riess, A.G. et al., preprint astro-ph/9805201, (1998).\\  
\noindent
[3] Ruiz-Lapuente, P., private communication (1999).\\
\noindent
[4] Felten, J.E., and Isaacman, R., {\it Rev. Mod. Phys.}, { \bf 58} (1986) 689. \\
\noindent
[5] Guth, A., {\it Phys. Rev. D}, {\bf 23} (1981) 347.\\
\noindent
[6] Carroll, S.M., Press, W.H., Turner, E.L., {\it Ann. Rev. Astron. \& Astrophys.}, {\bf 30} (1992) 499.\\
\noindent
[7] Ellis, G.F.R. et al., {\it Phys. Rep.}, {\bf 124} (1985) 315.\\
\noindent
[8] Krasi\'{n}ski, A., preprint gr-qc/9806039, (1998).\\
\noindent
[9] Ribeiro, M.B., {\it Astrophys. J.}, {\bf 441} (1995) 477.\\
\hspace*{-.1in}
[10] Mashhoon, B., Partovi, M.H., {\it Phys. Rev. D}, {\bf 20} (1979) 2455.\\
\hspace*{-.1in}
[11] Partovi M.H., Mashhoon, B., {\it Astrophys. J.}, {\bf 276} (1984) 4. \\
\hspace*{-.1in}
[12] Ellis, G.F.R., (1971) in {\sl General Relativity and Cosmology},
                ed. R.K. Sachs (N.Y.: Academic Press).\\
\hspace*{-.1in}
[13] Pascual-S\'anchez, J.F., preprint MAF-UVA-01, (1999).\\
\hspace*{-.1in}
[14] Kristian, J., Sachs, R.K. {\it Astrophys. J.}, {\bf 143} (1966) 379.\\

\end{document}